# Entanglement induced by Heisenberg exchange between an electron in a nested quantum dot and a qubit with relative motion

Li-Che Lin,[1] Seng Ghee Tan,[2] Ching-Ray Chang,[3] Shih-Jye Sun,[4] and Son-Hsien Chen[1,*]

[1]*Department of Applied Physics and Chemistry, University of Taipei, Taipei 10048, Taiwan*

[2]*Department of Optoelectric Physics, Chinese Culture University, Taipei 11114, Taiwan*

[3]*Quantum Information Center, Chung Yuan Christian University, Taoyuan, 320314, Taiwan*

[4]*Department of Applied Physics, National University of Kaohsiung, Kaohsiung 81148, Taiwan*

## Abstract

We propose a nested quantum dot (QD) structure to achieve enhanced control of entanglement induced by Heisenberg exchange between an electron and a qubit. Entanglement is quantified by mutual information (MI). We first consider the case of permanent exchange with the qubit embedded in a QD and find that the maximum MI (among all times) is solely determined by the initial angle between the spins. The entanglement exhibits distinct growth characteristics during different phases, and the saturation behavior mimics that of a strict zero-dimensional quantum dot, where exchange strength and interaction time combine into a single parameter, the amount of interaction. Introducing a nested dot with a moving qubit (decaying exchange) allows the electron to occupy either bound or scattering states, resulting in different final entanglement. In longitudinal motion, scattering states produce greater entanglement, while in transverse motion, the exchange range and qubit speed govern the entanglement behavior. The nested QD structure offers broad tunability of the final entanglement through gate voltage control, although the maximum achievable MI remains fundamentally limited by the maximum set in the case of permanent exchange. Our results demonstrate that combining system geometry, motion speed, and voltage control provides an effective strategy for optimizing entanglement tunability in QD devices.

## I. INTRODUCTION

Quantum entanglement represents a profound phenomenon in quantum mechanics, where two or more subsystems are intrinsically correlated such that the quantum state of each subsystem cannot be described independently [1, 2]. Comprehending how entanglement is built [3] is crucial for advancing both theoretical understanding and technological applications [2, 4–16] in quantum systems. One salient mechanism of generating entanglement is through the spin-spin exchange interaction of the form $\Gamma_{\alpha\beta} S_\alpha S_\beta$, where $S_\alpha$ and $S_\beta$ represent the spin operators along the $\alpha$ and $\beta$ directions ($\alpha$, $\beta \in \{x, y, z\}$), respectively. The Heisenberg exchange [17], contributes through the diagonal elements of the matrix $\Gamma$, while the Dzyaloshinsky-Moriya exchange [18, 19] appears through the off-diagonal elements; the Kaplan-Shekhtman-Entin-Wohlman-Aharony exchange [20, 21], manifests through both diagonal and off-diagonal elements [22, 23]. Among these exchanges, the Heisenberg exchange stands out, as it does not require spin-orbit interactions and demonstrates an entanglement phase transition in thermal states [24]. The critical temperature for this transition can be controlled through the exchange anisotropy and the external magnetic field [24, 25]. Furthermore, even weakly entangled states have been shown to be useful for achieving high-precision measurements [26]. The randomization of Heisenberg exchange [27] supports highly accurate magnetic field detection by measuring the state return probability [28].

* sonhsien@utaipei.edu.tw

The minimal set required to form entanglement is a two-qubit system. Particularly, the electron spin-qubit state in quantum dots (QDs) [29–32] has been experimentally realized, with the Heisenberg exchange interaction coupling between the qubits [33–35]. This minimal set exhibits exotic dynamics. For instance, when two qubits interact with a common environment, their entanglement can abruptly disappear [36–43] more easily than in systems beyond the minimal set. By applying an exponentially time-varying magnetic field, the final entanglement between the two qubits can be effectively preserved [44]. Additionally, in a closed two-qubit system, the evolved states span two separate Hilbert subspaces in the presence of a time-varying nonuniform magnetic field [45].

Moreover, entanglement becomes more amendable by permitting qubits to move freely in space. By tuning the velocities of two atom qubits moving within their respective cavities, which act as interacting local environments, the initial entanglement can be maintained [46]. For spin qubits, properly assigning the interaction time during which a qubit sequentially interacts with all others, a generalized W-state [47] is achievable [48]. In addition, when sending qubits to scatter off static qubits via local Heisenberg exchange, the transmission coefficients of the scattered qubits can be analyzed to yield tomography of the static spin qubits, enabling the reconstruction of their quantum states [49]. Robust entanglement between two ballistic electrons can be built from scattering off a magnetic impurity [50]. However, few studies have investigated how the interplay among confinement potential, exchange strength, interaction duration, and initial qubit configurations affects entanglement. It remains unclear which of these factors primarily determines the final entanglement.

In this paper, we examine the bipartite entanglement between an electron and a qubit with relative speed in the nonrelativistic regime. The qubit entangles with the electron via Heisenberg exchange. The electron is confined in a quasi zero-dimensional (0D) QD, namely of finite size in the $x$-direction.

For multiple qubits, entanglement can be established through interaction with a mediating spin. In the framework of the studied structure herein, this corresponds to using a series of moving qubits that are sequentially directed to interact with the QD. The resulting system resembles a spin-star network [51–55], with the interaction links activated one by one. If the resulting entanglement can be tuned over a wide range, a large entangled Hilbert space can be accessed. However, achieving such tunability requires understanding how to optimize the entanglement range between a qubit and an electron. Here, we first analyze the case of permanent exchange involving a local qubit embedded in a QD. We find that the exchange interaction duration and strength play a similar role in establishing the final entanglement, whose maximum value is pre-set by the initial spin angle between the electron and the qubit. Within this maximum, we then investigate how to optimize the tunable entanglement range by analyzing the case of a decaying exchange induced by a moving qubit. A nested quantum dot (NQD) is introduced to control the electron state, either bound or scattering, each exhibiting distinct entanglement behaviors. Our results demonstrate that gate voltage enables wide control over the entanglement range and clarifies the roles of the exchange range, qubit speed, and motional direction in gate-voltage-based tuning. Furthermore, the similarity between the mechanisms of interaction duration and strength offers a fresh perspective on describing and manipulating entanglement dynamics.

The paper is organized as follows. In Sec. II, we introduce the system under study and address the mutual information (MI) [56, 57], which serves as the entanglement monotone employed in this work. Our numerical simulations are presented and discussed in Sec. III for the QD (Sec. III A) and NQD (Sec. III B) systems. Section IV concludes our findings.

## II. MODELS AND METHODS

We consider the Hamiltonian that captures relative motion between the electron and the qubit as

$$\begin{aligned} H(t) &= \sum_i V_g(x_i)\, c_i^\dagger c_i - \gamma \sum_{\langle ij \rangle} c_i^\dagger c_j \\ &+ \sum_{i,ss's''s'''} J(x_i, t) \\ &\times c_{is}^\dagger c_{is'} d_{s''}^\dagger d_{s'''} \left(\vec{\sigma}^e\right)_{ss'} \cdot \left(\vec{\sigma}^q\right)_{s''s'''} . \end{aligned} \tag{1}$$

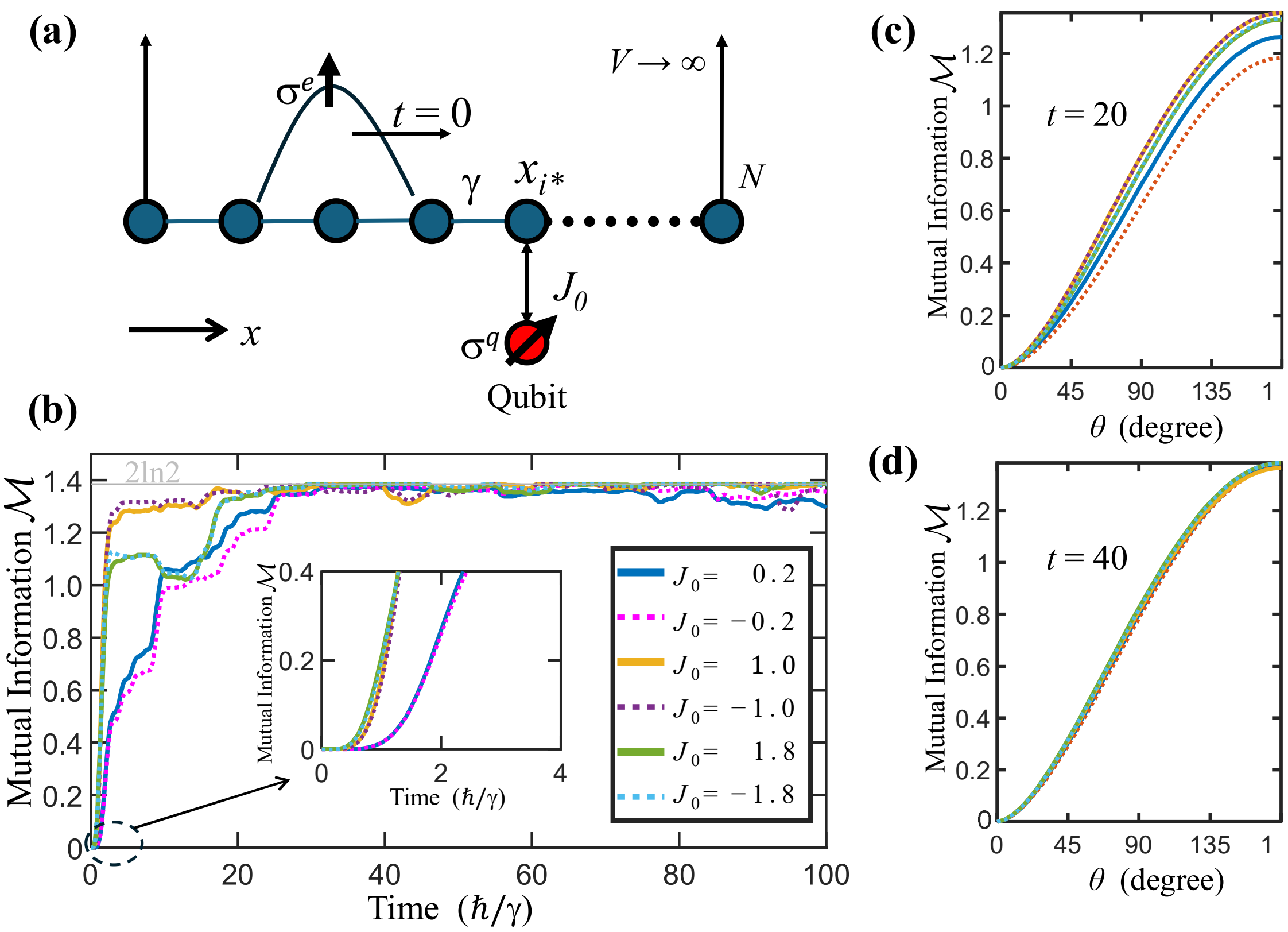

FIG. 1. (a) Schematic of a right-moving electron wave packet initially centered at $x_c = 12a$, which will interact, via Heisenberg exchange $J_0$, with the qubit of fixed position at $x_{i^*} = 16a$; $a$ is the lattice spacing. The quantum dot is quasi-zero-dimensional of finite size with $i = 1, 2, \cdots, N = 21$ lattice points formed by an infinite potential well ($V = \infty$ outside the dot). The initial angle between the electron spin $\vec{\sigma}^e$ and the qubit $\vec{\sigma}^q$ is $\theta$. (b) Mutual information at $\theta = \pi$ as a function of time $t$ for different values of $J_0$. The inset provides a zoomed view around $t \approx 0$. (c) and (d) mutual information as a function of $\theta$ at $t = 20$ and $t = 40$ for various values of $J_0$, respectively.

Here $c^\dagger_{is}(c_{js})$ denotes the creation (annihilation) operator that creates (annihilates) an electron with spin $s = \uparrow, \downarrow$ on site $i$. The operators $d^\dagger_s$ and $d_s$ represent the creation and annihilation of the qubit, respectively; the $\vec{\sigma}^e$ are the Pauli spin matrices of the electron, and $\vec{\sigma}^q$ the spin matrices of the qubit. The kinetic energy is characterized by the hopping $\gamma$ between two nearest-neighbor sites $\langle ij \rangle$. An infinite quantum well confines the electron within $x_{i=1}$ to $x_{i=N}$, as shown in Fig. 1(a). *Further* applying the gate voltage potential $V_g(x_i)$ produces a nested dot structure, as depicted in Figs. 2(a) and 3(a). The velocity of the moving qubit is introduced through the time-dependent exchange coupling $J(x_i, t)$.

Specifically, in the scenario of a qubit moving at constant velocity $\vec{v}^q$, Figs. 2(a) and 3(a), we assign

$$J(x_i, t) = \frac{J_0}{\sqrt{\pi} w_J} \exp\left(\frac{-\left|\vec{r} - \vec{v}^q (t - t_0)\right|^2}{w_J^2}\right). \tag{2}$$

The electron wave function $|\psi^e(x_i, t)\rangle$ varies with the position vector $\vec{r} = (x_i - x_0, y = 0)$ and is confined initially inside the NQD. The NQD centered at $x_0$ is subject to $V_g(x_i)$. The effective exchange distribution experienced by the electron is assumed to follow a Gaussian form, normalized by the prefactor $1/w_J$. This normalization allows entanglements in different interaction ranges $w_J$ to be compared on an equal footing. Furthermore, $w_J$ quantifies the spread of the interaction (i.e., the exchange interaction range) in units of the lattice spacing. We let the qubit pass over the dot center $x_0$ at time $t_0$, at which the maximum exchange takes place. On the other hand, in the scenario of an embedded qubit located at $x_{i^*}$ [Fig. 1(a)], realizable by a magnetic impurity, we adopt the local exchange,

$$J(x_i, t) = J_0 \delta_{i,i^*}. \tag{3}$$

The studied quasi-0D system can be carried out using gate-defined quantum dots [31]. Especially, exchange with the moving qubit is achievable in solid-state systems [58] such as a single-walled carbon nanotube [59].

The total wave function in the system evolves according to the Schrodinger equation,

$$i\frac{\partial}{\partial t} |\Psi(x_i, t)\rangle = H(t) |\Psi(x_i, t)\rangle, \tag{4}$$

(reduced Planck constant $\hbar \equiv 1$) with the separable initial conditions (ICs),

$$|\Psi(x_i, 0)\rangle = |\psi^e(x_i, 0)\rangle \otimes |\chi^e\rangle \otimes |\chi^q\rangle \tag{5}$$

where $|\chi^e\rangle$ and $|\chi^q\rangle$ are the spin states of the electron and qubit, respectively. The system is in a pure state described by the density matrix (DM) $\rho(x_i, x_j, t) = |\Psi(x_i, t)\rangle \langle \Psi(x_j, t)|$. The MI [56, 57]

$$\mathcal{M}(\rho) = \mathcal{S}(\rho^e) + \mathcal{S}(\rho^q) - \mathcal{S}(\rho) \tag{6}$$

detects quantum entanglements via the Von Neumann entropy $\mathcal{S}\left(\rho^{e/q}\right) = -Tr\left(\rho^{e/q} \ln \rho^{e/q}\right)$. Here $\rho^{e/q} = Tr_{q/e}\left(\rho^{e/q}\right)$ is the reduced DM of the corresponding subsystems. Since the system state is pure, $\mathcal{S}(\rho) = 0$, the two-particle (electron and qubit) correlations detected by the MI correspond to quantum correlations rather than classical correlations. We have also used the concurrence as the entanglement monotone and found that the identified features herein remain unchanged. To solve Eq. (4), each term in (1) is expressed as a $4N$ by $4N$ matrix using appropriate tensor products with identity matrices; for example, $\sigma_x^e \sigma_x^q$ becomes $I_{N\times N} \otimes \sigma_x^e \otimes \sigma_x^q$, where $I_{N\times N}$ is an $N$ by $N$ identity matrix. Throughout this work, we consider only pure states, as the system evolves under the time-dependent Schrodinger Eq. (1), which preserves the initially prepared purity $Tr(\rho^2) = 1$.

## III. NUMERICAL SIMULATION AND DISCUSSION

Regard the total *amount of interaction* characterized by,

$$\int_{-\infty}^{\infty} J^e(t)\, dt, \tag{7}$$

where $J^e(t) = \sum_{i=1}^{N} |\psi^e(x_i, t)|^2 J(x_i, t)$ quantifies the exchange electron experienced at time $t$. Three scenarios are considered. In the first scenario Fig. 1, the electron is injected into the QD, allowing for an *infinite* amount of interaction. In the second (third) scenario Fig. 2 (Fig. 3), the qubit moves longitudinally (transversely) through the NQD, generating a *finite* amount of interaction.

In what follows, we present our numerical results using (4) for a QD containing $N = 21$ sites. Exchange $J$ and potential $V_g$ energy are in units of $\gamma$, and time $t$ is in unit of $\hbar/\gamma$. The relative motion is modeled by either a moving electron or a moving qubit. Both cases employ an infinite quantum well. However, in the latter, a sub-well by $V_g$ confines the electron initially in its ground state, after which the electron is excited by the itinerant qubit that effectively provides a finite interaction as the exchange decays in the long-time limit. The former, on the other hand, produces a permanent exchange with the qubit that resides at a fixed position $x_{i^*}$ inside the dot.

### A. Electron injected into a quantum dot

Consider an electron injected from the left to the localized qubit, Fig. 1. We assume that the IC of the electron is given by a Gaussian wave packet,

$$\psi^e(x_i,0)=\frac{1}{\sqrt{C}}\exp\left(-(x_i-x_c)^2+ik(x_i-x_c)\right), \tag{8}$$

in Eq. (5). Here for right-moving electron prorogation, we set positive $k=1$, and $C$ is a constant assigned according to the probability normalization, $\sum_{i=1}^{N}|\psi^e(x_i,0)|=1$. The initial peak of the Gaussian $x_c=12$ of the electron and the qubit position $x_{i^*}=16$ (in unit of lattice spacing $a$) generating local exchange (3) are adopted; $\theta$ denotes the angle between the spins $|\chi^e\rangle$ and $|\chi^q\rangle$. Since the local exchange felt by the electron does not decay over time, the amount of interaction is infinite.

As shown in Fig. 1(b), rapidly growing entanglement is observed immediately after the electron wave function arrives at the site $x_{i^*}$. Following this growth phase, MI develops slowly and is modulated by the back-and-forth scattering of the electron bouncing off the infinite potential well. A certain maximum MI $\mathcal{M}_{\max}$ is then reached, and the system tends to build a saturation of MI eventually. In the early stage, as shown in Fig. 1(c), different exchange values $J_0$ yield different MI. However, as the system evolves further, Fig. 1(d), the MI follow a behavior that can be fit by $[1-\cos(\theta)]\times\ln 2$ and becomes insensitive to $J_0$ as well as the coupling type, whether ferromagnetic $J_0<0$ or antiferromagnetic $J_0>0$. These saturation ($t\gg 0$) features are similar to those of the strict 0D case $N=1$ [43], where exchange $J_0$ and time $t$ play the same role by entering the dynamics through a single variable $J_0\times t$ as Eq. (7) suggests. Interestingly, we observe that $\mathcal{M}_{\max}$ *depends only on the initial angle* $\theta$. In other words, the maximum entanglement is intrinsically pre-set (before interaction takes place) by $\mathcal{M}_{\max}(\theta)$; the initial angle becomes the sole relevant factor limiting the maximum possible entanglement that can be established, overriding interaction strength, dot size, and initial electron speed and distribution. In fact, we find that the same $\theta$-determined maximum $\mathcal{M}_{\max}(\theta)$ from the infinite-amount interaction also defines the maximum entanglement from the finite-amount interaction due to a moving qubit initially at the same $\theta$. Forms other than the Gaussian form of the initial wave packet can also result in the same maximum value, provided that the qubit and the electron remain in contact within a local spatial region [i.e., undergoing permanent exchange leading to infinite interaction in (7)]. With this understanding, we next discuss the scenario of a moving qubit.

### B. Qubit moving through a nested quantum dot

Consider now the qubit moving at a constant speed $v^q$. As aforementioned, the maximum is set by $\mathcal{M}_{\max}$. We choose the anti-parallel configuration $\theta = \pi$, for which $\mathcal{M}_{\max} = 2\ln 2 \approx 1.386$ [refer also to Fig. 1(d)]. According to Eq. (2), when the qubit moves longitudinally along the $x$-direction in Fig. 2(a), one has $\vec{v}^q = (v^q, 0)$ leading to

$$J\left(x_i, t\right) = \frac{J_0}{\sqrt{\pi} w_J} \exp\left(\frac{-\left[\left(x_i - x_0\right) + v^q\left(t - t_0\right)\right]^2}{w_J^2}\right), \quad (9)$$

in the Hamiltonian (1). When the qubit moves transversely along the $y$-direction in Fig. 3(a), one has $\vec{v}^q = (0, v^q)$ resulting in

$$\begin{aligned} J\left(x_i, t\right) &= \frac{J_0}{\sqrt{\pi} w_J} \exp\left(\frac{-\left(x_i - x_0\right)^2}{w_J^2}\right) \\ &\quad \times \exp\left(\frac{-\left(v^q\right)^2\left(t - t_0\right)^2}{w_J^2}\right). \end{aligned} \quad (10)$$

For a given $v^q$, the exchange (10) represents the case with the weakest amount of interaction among all scenarios studied herein. In the QD, an inner area is subject to a voltage potential $V_g\left(x_{i=8,9,\cdots 14}\right) = -0.4$, forming a NQD of size $N_0 = 7$. The selected $V_g$ produces 10 bound states in the NQD, and the electron is correspondingly prepared in the lowest energy state. We set exchange $J_0 = 5$, so that the electron initially trapped in the nested dot can be excited to the scattering state (i.e., outside the NQD) when the amount of interaction is enough. The qubit is initially assumed to be outside the dot and far from the electron by assigning a sufficiently large $t_0 = 1500$. At time $t_0$, the qubit arrives at the center of the NQD and then starts to move away from the dot.

The final MI $\mathcal{M}\left(t \to \infty\right)$ as a function of the interaction range $w_J$ and the qubit speed $v^q$ is shown in Fig. 2(b) for the longitudinal motion and in Fig. 3(b) for the transverse motion. The shaded region signifies the state where the electron is scattered outside the NQD. Note that the additional term $J_0/\left(w_J\sqrt{\pi}\right)\exp\left[-2v^q\left(x_i - x_0\right)\left(t - t_0\right)/w_J^2\right]$, as seen by comparing Eqs. (9) and (10), accounts for the longer interaction duration in the case of longitudinal motion. Therefore, for a given $v^q$, to achieve the same amount of interaction necessary for the electron to be excited to the scattering state, the transverse case requires a longer exchange range $w_J$. In the transverse case, but not the longitudinal case, our results indicate that when the qubit moves very slowly, the final MI alternates and exhibits high sensitivity to $w_J$. For a given $w_J$, in both cases, however, a smaller $v^q$ (and thus a longer duration) does not always lead to a larger saturated MI. In fact, the saturated MI oscillates with $v^q$. In the limit of very fast speed $v^q$ and very short range $w_J$, MI decays to zero due to extremely small interaction duration. Nonetheless, a feature independent of the exchange range $w_J$ and the qubit speed $v^q$ is that, for longitudinal motion, the scattering states generally yield larger saturated entanglement than the bound states.

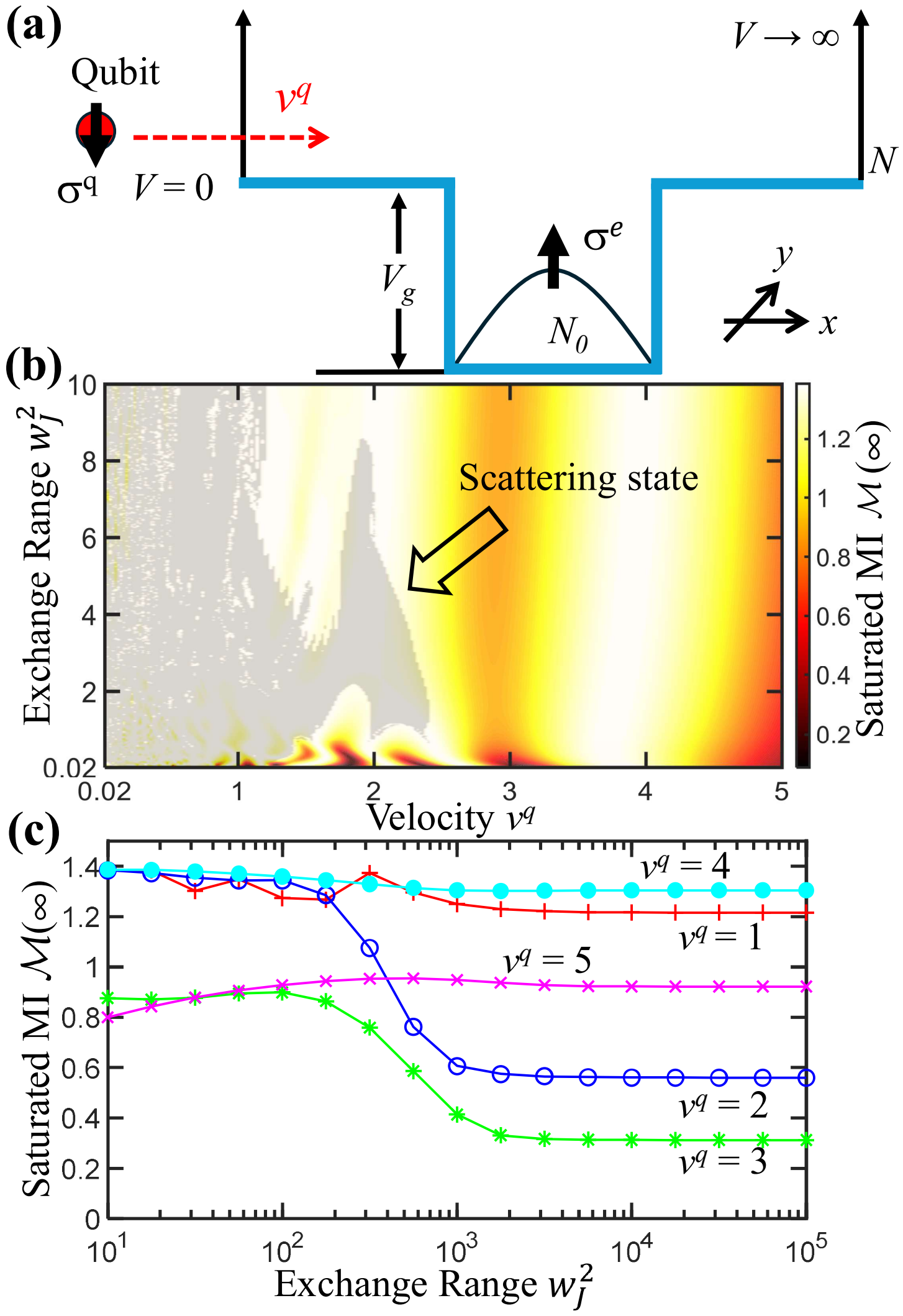

FIG. 2. (a) Schematic of a qubit moving longitudinally along the $x$-direction at a constant speed $v^q$ through a nested quantum dot of size $N_0$ and applied gate voltage potential $V_g$. The electron is prepared in the ground state, confined within the nested dot, and experiences the effective exchange Eq. (9) due to the moving qubit. (b) Saturated mutual information $\mathcal{M}(t \to \infty)$ as a function of the exchange range quantified by $w_J^2$ and speed $v^q$. The shaded region indicates where the electron becomes excited out of the nested dot, transitioning to a scattering state. (c) Logarithmic plot of the $w_J^2$ dependence of the saturated mutual information at various $v^q$.

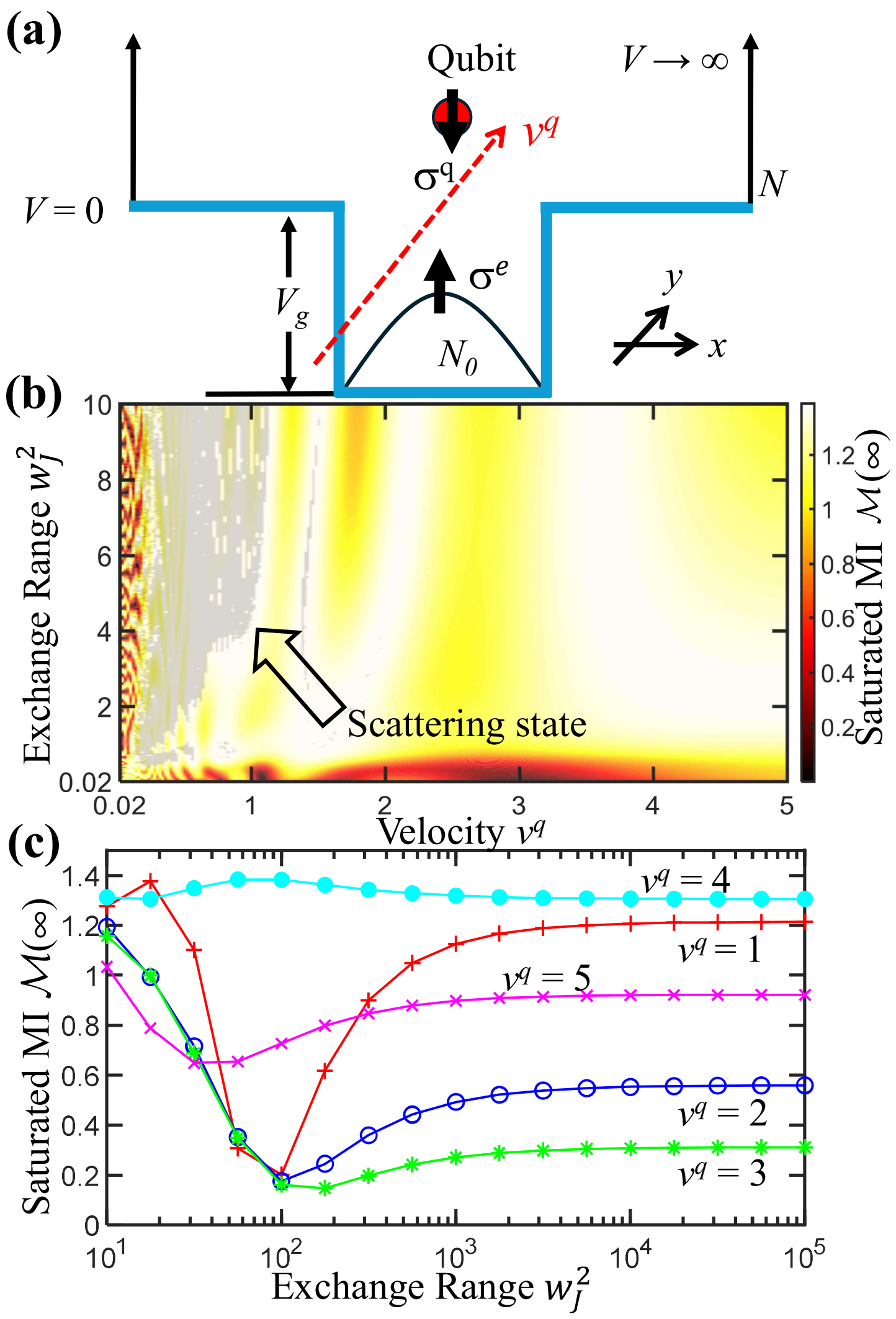

FIG. 3. (a) Schematic of the nested quantum dot similar to Fig. 2 but with the qubit moving transversely along $y$-direction, described by the exchange, Eq. (10). (b) Saturated mutual information $\mathcal{M}(t \to \infty)$ as a function of $w_J^2$ and speed $v^q$. The shaded region indicates the scattering states. (c) Saturated $\mathcal{M}(t \to \infty)$ as a function of the logarithmic scale of $w_J^2$ for different values of $v^q$. Note that this represents the scenario with the weakest amount of electron-qubit interaction among all the device structures proposed herein.

The benefit to introduce the NQD then emerges. The device is multi-functional, allowing both scattering and bound states to be utilized for different purposes. If larger entanglement is desired, the scattering state is preferable. If tunability of the final entanglement is required, bound states can be exploited by adjusting the qubit speed $v^q$ at moderate values.

Figures 2(c) and 3(c) illustrate how the final MI $\mathcal{M}\,(t \to \infty)$ varies with increasing $w_J$ for longitudinal and transverse motion, respectively. We note that the qubit eventually interacts effectively with a single-electron QD of zero size due to their large separation distance, regardless of the moving direction of $\vec{v}^q$. However, for mild $w_J \lesssim N$, the final entanglement retains a memory of the $\vec{v}^q$ direction and finite NQD size experienced around $t \approx t_0$. This can be seen by comparing $\mathcal{M}\,(t \to \infty)$ in Fig. 2(c) and in Fig. 3(c) for $w_J^2$ less than $10^3$. Conversely, for long-range interaction, $w_J \gg N$, the memory effect of the moving direction is lost, i.e., both longitudinal and transverse $\vec{v}^q$ yield the same $\mathcal{M}\,(t \to \infty)$. This loss suggests that the spatial information of the qubit (or spatial dependence of the exchange) is absent. Being worth noticing, even in the uniform limit of the exchange distribution, which leads to very weak but long-lasting interaction, significant entanglement can still be established, as indicated by the converged $\mathcal{M}$ as $w_J \to \infty$.

Figures 4(a) and 4(b) present the saturated $\mathcal{M}\,(t \to \infty)$ at an initial spin angle $\theta = 45^\circ$ for longitudinal and transverse qubit motion, respectively. The entanglement behavior remains consistent, with the maximum value pre-set by the initial angle $\theta$ [see 1(d), where $\theta = 45^\circ$ yields $\mathcal{M}_{\max} \approx 0.32$]. In longitudinal motion, the scattering state produces higher entanglement, while in transverse motion, slow qubit speeds make the final mutual information highly sensitive to the exchange range $w_J$. Note, however, that this sensitivity appears only after the qubit has long passed the NQD and does not arise near $t \approx 0$. The above features generally hold for any given initial angle $\theta$.

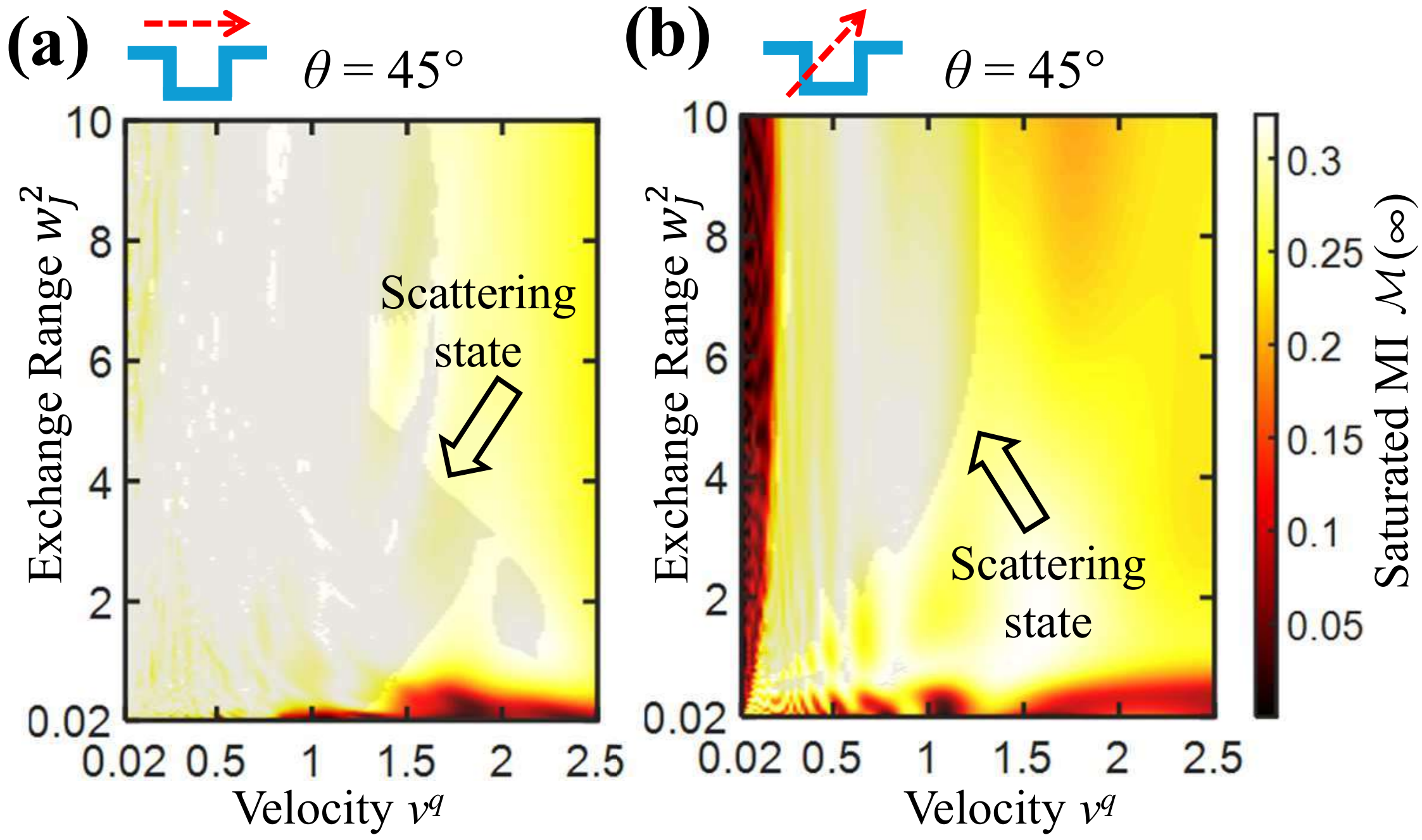

FIG. 4. Saturated mutual information $\mathcal{M}(t \to \infty)$ as a function of $w_J^2$ and qubit speed $v^q$ at $\theta = 45°$ for (a) longitudinal and (b) transverse qubit motion.

Although $\theta$ sets the main limit on the final MI, the role of gate voltage in expanding the entanglement range remains unclear. Here, we address this point. To examine the tunability of entanglement through the nested quantum dot, $\mathcal{M}(t \to \infty)$ is plotted as a function of gate voltage $V_g$ for various speeds $v^q$ and interaction ranges $w_J$, as shown in Fig. 5(a) for longitudinal motion and Fig. 5(b) for transverse motion. Several important features emerge. First, the slower speed $v^q = 0.02$ yields a wider range of final MI compared to the higher speed $v^q = 1$. Second, at $v^q = 0.02$, increasing $V_g$ from a small value reveals an oscillatory-like dependence, whose period reflects the interaction range $w_J$. Specifically, larger $w_J$ leads to faster oscillations in $\mathcal{M}(t \to \infty)$. Finally, transverse motion results in greater variability in the MI compared to longitudinal motion. In general, a broader entangled state Hilbert space can be accessed through gate voltage control at smaller $v^q$, with $w_J$ determining how rapidly the final entanglement oscillates with $V_g$; this behavior has been verified for different initial spin angles $\theta$ (not shown here).

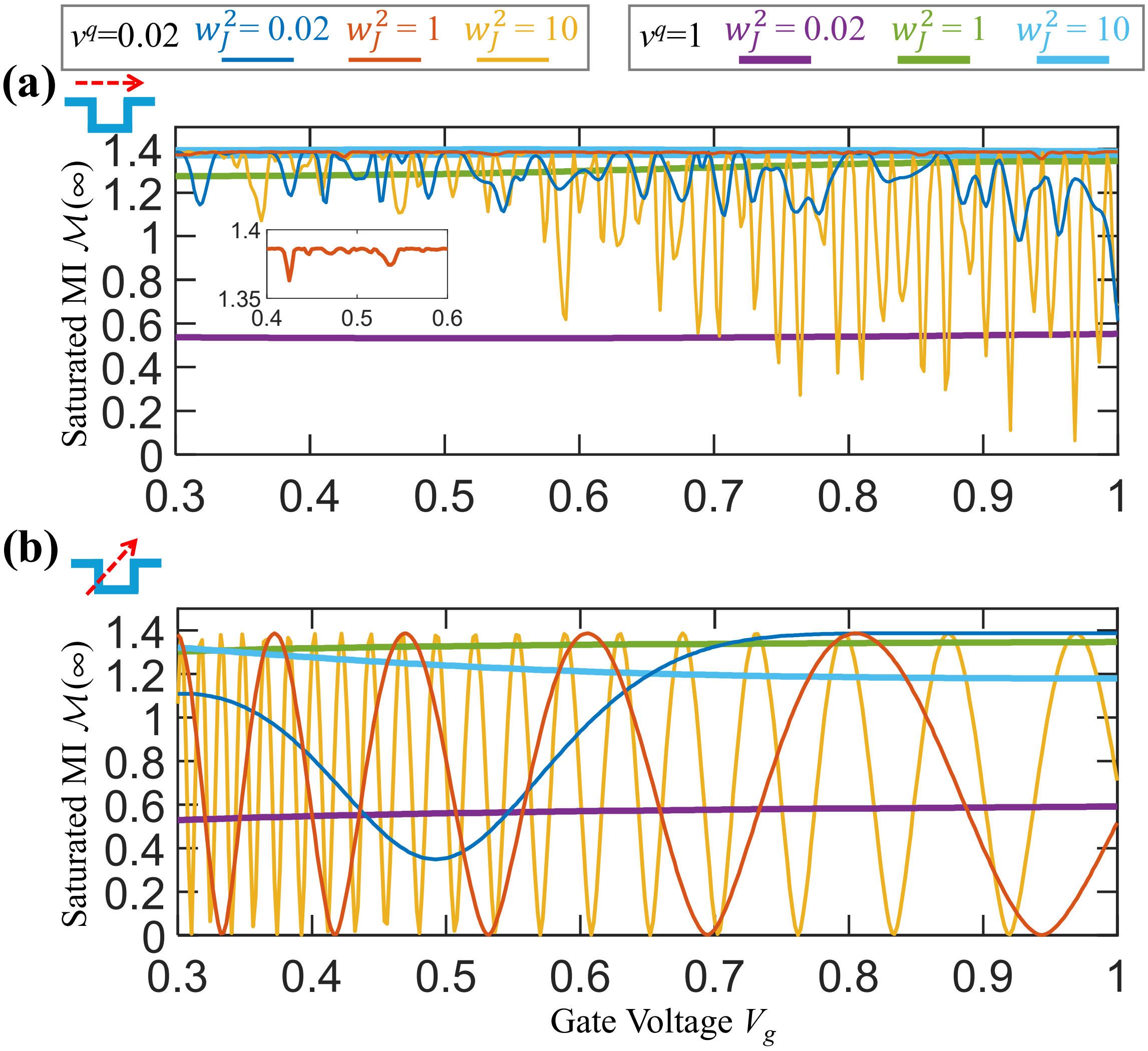

FIG. 5. Saturated mutual information $\mathcal{M}(t \to \infty)$ at $\theta = 180°$ as a function of gate voltage $V_g$ for (a) longitudinal and (b) transverse qubit motion, evaluated at qubit speeds $v^q = 0.02$ (thin lines) and $v^q = 1$ (thick lines) for various $w_J$. The inset in (a) shows a zoomed-in view for $w_J = 1$ at the slower speed $v^q = 0.02$.

We analyze the realizability of the qubit speeds used in our simulation. Recall that time $t$ is measured in units of $\hbar/\gamma$, so the speed $v^q$ is expressed in units of $a\gamma/\hbar$. Taking a lattice spacing of $a \approx 0.5$ nm and adopting hopping energies $\gamma$ between 0.1 eV and 1 eV, which are typical for tight-binding atomic chains, the speed range from $v^q = 0.02$ to $v^q = 1$ maps to physical speeds between approximately $1.5 \times 10^3$ m/s and $7.6 \times 10^5$ m/s. These values are consistent with typical Fermi velocities in GaAs-based two-dimensional electron gases (approximately $10^5$ m/s) [60] and in graphene (approximately $10^6$ m/s) [61, 62]. They are also comparable to charge mobilities corresponding to low carrier velocities in molecular chains (approximately $10^3$ m/s) [63, 64]. Therefore, the speeds employed in our study are feasible and lie within experimentally accessible regimes in these nanodevices. In particular, the motion of the qubit in our present study, when implemented with a single electron, can be realized using electron shuttling techniques [65–67] and single-electron pumps [68–71] to achieve even slower speeds, thereby allowing access to a larger entangled state space.

## IV. CONCLUSION

In conclusion, we explore the entanglement behavior between an electron confined in an infinite quantum well (QD) and a qubit with relative motion. The Heisenberg exchange between the two spins induces entanglement. When the qubit is embedded in the dot, a moving electron described by an initial Gaussian wave packet yields permanent interaction. In this case, the total amount of interaction (7) is infinite, and the maximum (among all times) MI $\mathcal{M}_{\max}$, is pre-set solely by the initial spin angle $\theta$. A rapid growth phase of entanglement is observed as the electron wave function reaches the site coupled to the qubit. This is followed by a slower growth during the near-saturation phase, where the electron undergoes back-and-forth scattering within the well. The saturation (set also by $\mathcal{M}_{\max}$) being independent of the exchange, resembles that of the strict 0D QD, where the amount $\int_{-\infty}^{t} J^e(t')\, dt'$ simplifies to $J_0 \times t$, determining the system evolution and effectively equating the roles of the exchange and time.

The same $\mathcal{M}_{\max}(\theta)$ also governs the maximum MI in the case of a finite amount of interaction, considering a moving qubit passing through the NQD, with the initial electron in the ground state. The nested structure enables versatile control of the final entanglement. When the qubit moves slowly and the exchange strength is large, the electron can be excited into scattering states due to a sufficient amount of interaction. Particularly for longitudinal motion, these scattering states enable the generation of large entanglement. In contrast, in the moderate- to high-speed regime, where the electron remains in the bound state, the qubit speed can be used to adjust the final entanglement. For short-range exchange interactions, longitudinal and transverse motion lead to different final entanglement. Interestingly, for long-range exchange interactions, the final entanglement becomes insensitive to the moving directions. Even with a uniformly weak exchange distribution, substantial MI can still be obtained. Our findings across all scenarios suggest that weak but long-lasting exchange interactions can yield a considerable amount of entanglement, with the primary bottleneck limiting the entanglement being the initial angle between the two spins.

The gate voltage applied to the proposed NQD provides broad control over the final entanglement, particularly in the regime of slow relative motion, where the interaction range induces a periodic or oscillatory dependence of entanglement on gate voltage. Consequently, the NQD provides a promising platform for accessing a large entangled state space and supporting multi-qubit state generation through spin-star networks, with the QD acting as the central spin. The identified features deepen our understanding of how the initial angle, interaction duration, interaction range, qubit speed, and motion direction govern Heisenberg-exchange-based entanglement. In light of these results, the findings highlight the potential of the NQD architecture as a versatile scheme for generating and tuning entanglement, offering valuable applications in quantum sensing, spin-network engineering, and solid-state quantum information processing.

## ACKNOWLEDGMENTS

One of the authors, S.-H. C., expresses gratitude to Ming-Chien Hsu and Che-Chun Huang for their valuable discussions. S.-G. Tan and C.-R. Chang acknowledge support from the National Science and Technology Council of Taiwan under Grant Nos. NSTC 113-2112-M-034-002 and NSTC 113-2112-M-033-011, respectively.